# Phosphorous Diffusion Gettering of Trapping Centers in Upgraded Metallurgical-Grade Solar Silicon

*Sergio Catalán-Gómez\*, Nerea Dasilva-Villanueva, David Fuertes Marrón, Carlos del Cañizo*

Instituto de Energía Solar, Universidad Politécnica de Madrid, ETSI Telecomunicación, Av. Complutense 30, E-28040 Madrid (Spain).

**Abstract**

In this paper we show experimental evidence indicating the beneficial impact of a phosphorous diffusion gettering (PDG) in the reduction of trapping centers, as observed by means of inductively-coupled photoconductance decay and lifetime measurements performed on upgraded metallurgical grade silicon (UMG-Si) wafers. We have detected the presence of trapping species dominating the long time range of the photoconductance decay of UMG-material (slow traps), which can be effectively removed after a PDG carried out at 780 ºC. Notwithstanding, a second trapping mechanism, characterized by a shorter time constant, still governs the response at very low injection levels after the gettering. Furthermore, the beneficial effect of the PDG is studied as a function of processing time, showing minority carrier bulk lifetime improvements up to 18-fold, up to the range of 70 μs. This work paves the way for developing gettering strategies capable of successfully removing trap centers and improving the bulk lifetime of unconventional Si material.

The photovoltaic industry is currently mostly based on polysilicon feedstock purified by the Siemens process, which is expensive in terms of both economic and environmental cost. To overcome this issue, alternative approaches, such as metallurgical purification routes, have been widely investigated. Specifically, upgraded metallurgical-grade silicon (UMG-Si) has gained attention from the scientific community over the last years and has become an alternative feedstock source for different companies[1–3] in the Si photovoltaic industry. UMG-Si offers not only lower cost than conventional polysilicon (whereby the cost of the final PV module can be reduced by about 8%), but also a reduced energy payback time (by 50%) and environmental impact, due to its reduced-emission purification process (quantified in as much as 70% cut of green-house-gas emissions, depending on the energy mix at the fabrication site)[3]. Despite the lower purity of the resulting material, containing a significantly high impurity concentration, it has been reported that multicrystalline solar cells based on UMG-Si show a negligible penalty in terms of efficiency (with values up to 20.76%) in Al-BSF and PERC architectures, as compared to standard polysilicon-based cells[3–5] and hence it can compete with conventional Si.

Lifetime measurements[6,7] are a common way of characterization of all types of Si wafers, and is of particular importance in our work. Carrier lifetime, defined as the characteristic time it takes the excess minority carrier to recombine, is mainly affected by four mechanisms: band-to-band recombination, negligible in Si; Auger recombination, surface recombination and bulk Shockley-Read-Hall (SRH) recombination, the latter proportional to the amount of recombination centers present in the bulk of the material, and therefore, the most critical mechanism, particularly in UMG-Si.

Different characterization techniques to measure carrier lifetime have been proposed, being photoconductance (PC) decay in its various forms the most popular one. For materials with low minority carrier lifetime the method used is the quasi-steady-state photoconductance (QSSPC)[8]. Working with this technique, an anomalous characteristic behavior is often observed, particularly at low injection levels[9]. Although the PC decay after the light pulse does not generally follow simply a single exponential function, the multi-exponential character of the curves does imply the existence of more than a single electronic process involved in the PC-decay. Such concurrent processes, to be ascribed to trapping centers, result in prolonged apparent lifetimes, that increase markedly as the excess carrier density decreases. These apparent lifetimes can reach values in the range of hundreds of microseconds or longer, a phenomenon typically associated to trapping effects and which has been observed in different types of Si (single crystal, polysilicon, thin film)[10].

Most of the previous works agreed to treat traps as electronic centers whose main characteristic is a significant difference in the values of the capture cross-sections of minority and majority carriers, whereas in the case of very effective recombination centers those values are typically of a similar order of magnitude[11]. Indeed, the Shockley-Read-Hall (SRH) formalism has been adapted to treat traps as recombination centers[12], and it has been observed experimentally, when comparing with other characterization techniques, that trapping centers can increment the recombination at low injection levels[13].

Along this line, in this work we report on the net reduction of bulk recombination in UMG Si wafers resulting from the removal of active defects and trapping centers upon a phosphorous diffusion gettering (PDG) step.

There have been just a few works that reported on a beneficial effect regarding the suppression and/or reduction of trapping effects in multicrystalline Si (mc-Si) after physico-chemical treatments. For instance, S. Jafari *et al.*[14] achieved a full removal of trapping effects in mc-Si by applying an infrared laser treatment; however, the bulk lifetime values of treated wafers were also affected and considerably reduced due to a higher recombination rate. On the other hand, Macdonald *et al.*[15] showed that the PDG treatment can have a two-fold effect: an increase of the minority carrier lifetime and a moderate reduction of the trap concentration. The latter approach is of great interest for UMG material, because of its low starting lifetime values.

Therefore, with the aim of studying the removal of trap centers, we have performed PDG processes on wafers grown from UMG-Si feedstock. The fabrication details of the UMG-Si material can be found elsewhere[3], although it has to be noted that the wafers used in this study correspond to a highly-contaminated batch, while the solar cell results in [3] were obtained from wafers coming from an optimized purification process. We have worked with sixteen UMG-Si bare wafers of 15x15 cm$^2$ size, 148 ± 2 μm thickness and 1.05 ± 0.07 Ω·cm average resistivity. The wafers were analyzed with the Sinton Instruments WCT-120 lifetime tester. The minority carrier lifetime is obtained from the values of the PC, using the Dannhäuser mobility model and a optical constant of 0.7. It is represented as a function of the excess carrier density in Figure 1 (a). As a result of the high impurity concentration, the recombination rate is very high in this type of Si and hence its bulk lifetime is also very low, in the range of 1 μs for the case of non-passivated surfaces. As it can be observed, at low injection levels, trapping, revealed as a large increase in the apparent lifetime (with values above 10 μs), is the dominant process. Depletion-region modulation (DRM) can be discarded as the main cause of the lifetime increase, since there are no p-n junctions in the bare wafers[9,12].

Hornbeck and Haynes[16] postulated models for the quantification of the trap parameters and later on Macdonald[17] adopted one of them to QSS conditions. In this work we have used an alternative strategy previously reported in the literature[14,18]. It is based on the analysis of the PC raw measurements which allow to distinguish different trap species using the PC decay time constant rather than characterizing the trap properties.

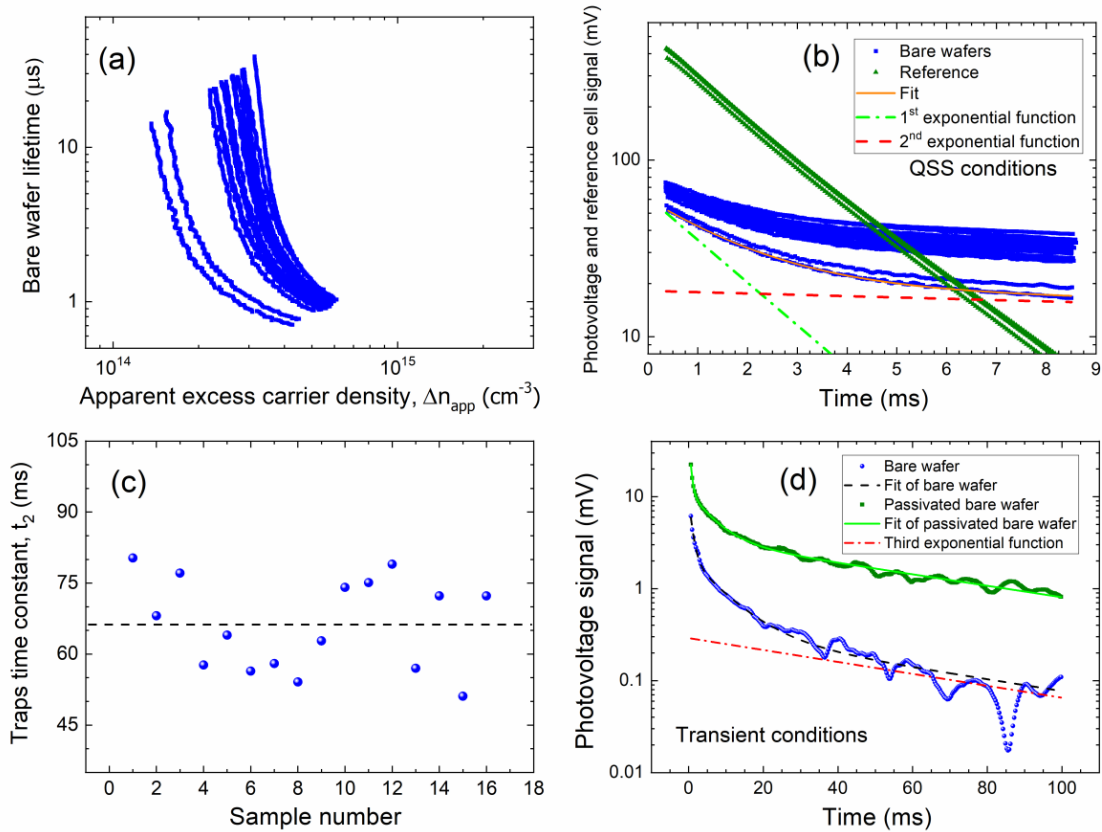

**Figure 1.** Sinton Instrument measurements: (a) Apparent lifetime of sixteen bare wafers as a function of the apparent excess carrier density. (b) Photovoltage signal of the wafers (blue squares) and the output voltage of the reference cell (olive triangles) versus time. Green dot-

dashed and red dashed lines represent two exponential functions used to fit (solid orange line) an exemplary curve. (c) Trap decay time constant ($t_2$) extracted from the second exponential function of all wafers. Horizontal dashed line represents the average value of all points. (d) Photovoltage signal of a bare wafer and a passivated one versus time under transient conditions. The corresponding fits of each curve are also included.

Figure 1 (b) shows the photovoltage signal, which is proportional to the photoconductance, of the analyzed wafers, together with the corresponding signal measured at the reference cell of the instrument (green dots), as a function of time. The slow flash mode used has a decay time constant of 1.8 ms, which is standard for QSS conditions used to measure samples with lifetime below 200 µs. The samples of this study have a high recombination rate resulting in low excess carrier densities upon illumination. This leads to a weak dependency of the lifetime with the injection level and thus, the corresponding PC measurement should follow the same exponential trend of the flash, with a similar decay time. However, the blue squares clearly deviate from linearity (mind the log-scale). This deviation is well understood as the result of trapping, as explained elsewhere[14,18]. The PC measurments could be satisfactorily fitted with a two-exponential function of the form $y=A_1·exp(-x/t_1)+A_2·exp(-x/t_2)$, with characteristic decay times $t_1$ and $t_2$. The first exponential function represents physical processes that happen in the time lapse of the flash, such as SRH-recombination of the measured sample ($t_1 \sim t_{flash}$), while the second one represents events that last longer, such as the ones governed by slow traps. As an example, a PC measurement is fitted and their two different exponential functions are plotted in Figure 1 (b). For this specific PC curve, the decay time constants ($t_1$ and $t_2$) obtained were 1.7 and 74 ms, respectively. Figure 1 (c) represents the slow trap decay time constants ($t_2$) obtained from the fitting of all sixteen PC measurements. All samples show $t_2$ values much higher than the flash decay constant, taking in average 66 ± 9 ms to decay (marked as a horizontal dashed line in the figure). All bare wafers have been considered to be dominated by the same slow trap center. However, it is important to notice that, although the PC curves can be properly fitted with two exponential functions, the presence of additional fast traps cannot be excluded at this stage.

For the fitting of the PC measurements we have followed the model previously described by Zhu *et al*[18], which considers that the long-term related PC decay time constant is only impacted by the trap parameters and therefore it can be used as a signature of the trapping center. However, this model relies on two important assumptions: (a) the trap decay is measured in transient condition; and (b) the re-capture of minority carriers in the trap is negligible. With the aim to check the first assumption, Figure 1 (d) shows the PC measurement of a bare wafer (blue spheres) with a short flash mode ($t_1$~0.02 ms), common of Sinton's transient mode. It can be seen how the PC signal is recorded for a longer time than that under QSS conditions. The corresponding fit (black dashed line) is also plotted. A total of three exponential functions are needed for an appropiate fit of this curve, being the decay time constant of the slowest exponential function equal to 69 ms. The agreement of this value with the one obtained under QSS conditions (Figure 1 (b)) demonstrates that the transient condition is fulfilled. Figure 1(d) also includes a PC measurement of a passivated bare wafer (olive squares) and its corresponding fit curve (green solid line). The initial PC value of this sample is higher compared with the unpassivated bare wafer due to its lower recombination rate (higher lifetime). The exponential function associated to the trapping center is almost the same (decay constant of 71 ms) for the passivated one, demonstrating that the lifetime value does not affect the trap parameters. Thus, the re-capture of minority carriers is negligible, as required by Zhu´s et al[18] model.

The samples were subsequently subjected to a PDG treatment. It is a well-known method to eliminate a high range of interstitial impurities in Si material, particularly effective with metallic species, and its bases are well stablished and understood[19,20]. After carefully cleaning the samples by CP4 (HNO$_3$/HF) and RCA1 chemical treatments, the samples were introduced in a tubular furnace at 780 ºC under N$_2$ and O$_2$ gas fluxes for the subsequent PDG from a POCl$_3$ liquid source. The process time has been varied from 30 to 120 min in 4 batches of 4 samples each followed by a 10 min drive-in providing an n-type emitter atop the Si wafer surface. The emitter is subsequently removed with CP4 etching and the wafers are passivated with 0.1 M iodine-ethanol[21] to minimize surface recombination effects. Recombination lifetime is measured after each step of the process, namely: before PDG, with and without passivation (already shown in Figure 1 (a)), with an emitter in both sides of the wafer, and after emitter removal, with and without passivation. Iodine-ethanol passivation avoids other derived consequences on trapping, such as those related to hydrogen incorporation or additional temperature effects, as it could happen with passivation layers based on SiN$_x$:H. For the sake of clarity, the passivated lifetime curves of the four wafers of the 90 min PDG batch are presented in Figure 2 (a). There is a clear difference in lifetime values before (green spheres) and after (red stars) the PDG. The gettering process successfully works by removing recombination impurities, thereby increasing the lifetime from 3.5-5 µs to 18-26 µs, as evaluated at $10^{15}$ cm$^{-3}$ injection level. Moreover, it is worth noting that the PDG does not completely suppress the trapping, since it is still present at low injection levels. Figure 2 (b) represents the lifetime values before and after the gettering as a function of the process time.

Before PDG the sixteen passivated wafers (green spheres) show a lifetime average value of 4 ± 2 µs. After gettering, the lifetime values depend highly on the process duration. The best results in terms of lifetime are found for the shortest gettering process (30 min), obtaining values from 42 to 73 µs, which means an improvement of 10 to 18 times the average lifetime before gettering. As observed in Figure 2 (b), for PDG durations higher than 30 min, the lifetime shows a lower improvement. This fact contrasts with other works on UMG Si where lifetime increases with increasing gettering times[22]. We can tentatively ascribe the observations to an effective activation of recombination paths triggered thermally. Due to the relatively low gettering temperatures used in our study, the activation can result in a metastable configuration of the responsible impurities with a cumulative character in the range of explored processing times. Incomplete internal gettering[23] cannot be ruled out as a possible explanation of the trend in Figure 2 (b). The variations of lifetime values between samples of the same batch are likely due to intrinsic factors such as different grain size, grain boundaries, dislocation densities or impurity concentration, as typically reported in mc-Si[24].

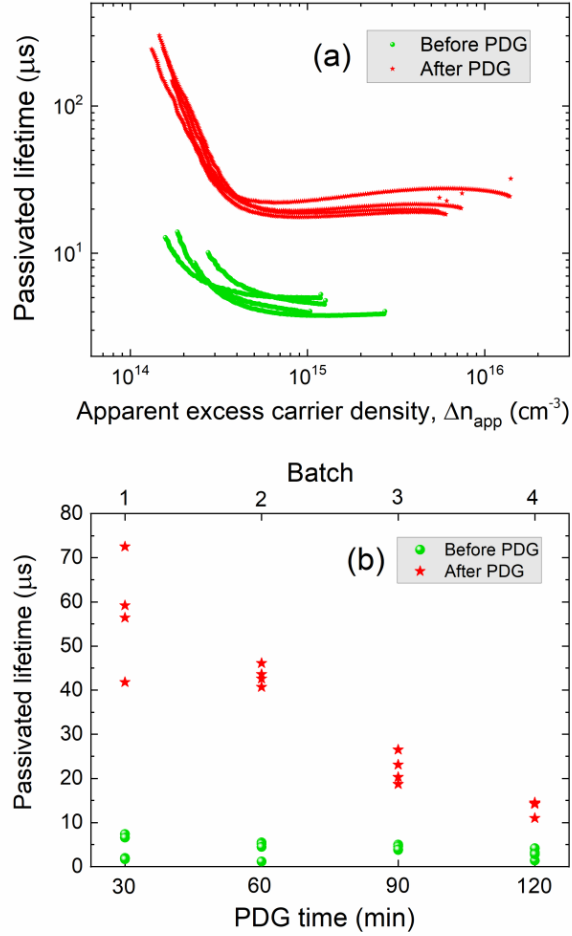

**Figure 2.** (a) Lifetime measurements of passivated samples before (green spheres) and after (red stars) PDG of the wafers of a representative batch after 90 min gettering process. (b) Lifetime values before and after PDG of all the batches processed at different PDG times evaluated at $10^{15}$ cm$^{-3}$ injection level.

From the lifetime curves of Figure 2 (a), changes in trapping effects due to the gettering process cannot be easily observed. To do so, we have proceeded in the same way as in Figure 1 (b). We have analyzed the PC spectra of the samples after PDG and compared them with pre-gettering data. Results are presented in Figure 3 (a) for an exemplary case of a PDG time of 30 min. Two major changes are observed after the gettering process. Firstly, the PC signal is increased at short times, which means lower overall recombination, in agreement with the lifetime improvement shown in Figure 2. Secondly, the PC curvature at longer times decreases after gettering. The latter is related with the decay time constant of the second exponential ascribed to the slow trap center. Figure 2 (b) shows this parameter ($t_2$) for all the samples before and after the PDG. As we concluded from Figure 1, all as-grown samples are dominated by the same slow trap center, whose decay time constant was estimated around 67 ms (blue spheres). However, after the PDG process, $t_2$ decreases to an average value of 14 ± 7 ms (still significantly larger than the 1.7 ms of the reference decay), thus revealing a change of the main trapping mechanism.

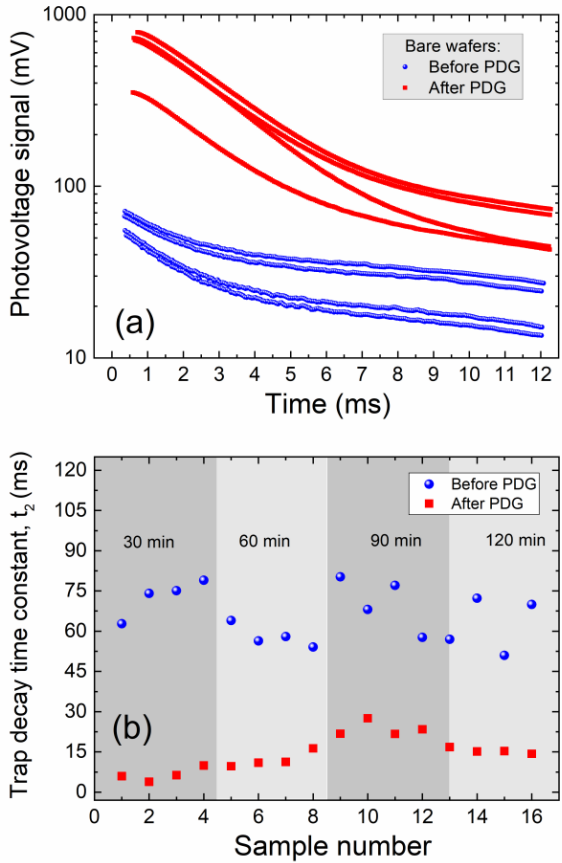

**Figure 3.** (a) Photovoltage signal of bare wafers of a representative batch before (blue spheres) and after (red squares) 30 min PDG. (b) $t_2$ obtained from the fitting of the PC

measurements before and after PDG as a function of the sample number. The PDG time is indicated in the figure.

In a few words, despite the PDG process clearly removes one type of slow trap, one additional trapping mechanism still remains active in the samples. Interestingly, Macdonald *et al*[15] reported two types of traps in mc-Si, one related to dislocations, impervious to gettering; and a second one which was getterable. This scenario seems to be also valid for contaminated UMG-Si, according to Figure 3. Although the presence of an interstitial metallic impurity acting with a dual behavior, as a trap and as a recombination center, have been reported previously[25], we cannot exclude that the electronic transport of the samples is actually governed by various active defects.

In summary, it has been shown that a phosphorous diffusion gettering process has a significant impact on the trapping of contaminated UMG-Si wafers. Traps controlling the long photoconductance decay time constant have been observed in as-grown samples. After PDG, those trapping centers disappear, revealing the presence of fast traps associated with a shorter PC decay time constant. The impact of the PDG process on trapping adds up to its well-known beneficial effect on the reduction of active recombination centers, resulting in an 18-fold increase of the lifetime values, and thereby reaching dozens of µs. The mechanisms unveiled in this work can help understand potential side effects of PDG on the reduction of recombination in mc-Si produced by metallic impurities acting as traps and recombination centers.

**Acknowledgements**

The Spanish Agencia Estatal de Investigación is acknowledged for funding through the SOLAR-ERA.NET Cofund project "Low Cost High Efficient and Reliable UMG PV cells

(CHEER-UP)", PCI2019-111834-2/AEI/10.13039/501100011033. The Comunidad de Madrid is also acknowledged for support under the project Madrid-PV2 (S2018/EMT-4308) Juan José Torres is acknowledged for support in wafer processing and, together with Ignacio Tobías, Bo-Kyung Hong, and Manuel Funes, also for fruitful discussions. Aurinka PV is acknowledged for wafer supply.

Received:

Revised:

Published online: